\documentclass[aps,prl,preprint,superscriptaddress,showpacs]{revtex4}

\usepackage{amssymb}

\usepackage{epsfig}

\usepackage{amsmath}

\usepackage{times}

\begin{document}

\title{Spiral Wave Solutions of One-dimensional Ginzburg-Landau Equation By Extended F-expansion Method}

\author{Xurong Chen}
\email{chen@physics.sc.edu}

\affiliation{Physics Department, University of South Carolina, Columbia, SC29208, USA}

\begin{abstract}
The one-dimensional Ginzburg-Landau (GL) Equation is considered. We use the recently developed
extended F-expansion method to obtain spiral wave solution of one-dimensional GL Equation.
\end{abstract}

\date{\today}



\maketitle

\section{Introduction}

The Ginzburg-Landau (GL) Equation is one of the most-studied nonlinear partial
differential equations (PDE) \cite{rmp2002}. Spiral waves are important 
patterns in many systems, such as physics, chemistry, materials, biology, etc.
Exploring exact solutions of nonlinear PDE is a hot and difficult topic in mathematical
physics. Many methods were developed for studying exact solutions of nonlinear
PDEs. Recently, a new method named extended F-expansion was proposed \cite{yangkq2004}\cite{wangml2006}.
It's useful to obtain more Jacobi elliptic function solutions. In this paper 
it is used to study the one-dimensional GL Equation. We obtain a spiral wave solution, 
which is exactly same with Hagan's solution \cite{hagan1982}.

\section{Spiral Waves of One-dimensional Ginzburg-Landau Equation}

The Ginzburg-Landau Equation is given by
\begin{equation}\label{GLequ}
\partial_{t}A = A + (1 + ib)\Delta A -(1+ic)|A|^2A, 
\end{equation}
where $A$ is a complex function of time $t$ and space $x$; $b$ and $c$ are
real parameters describing linear and nonlinear dispersion, respectively.
It's useful to represent the complex function $A$ in the following way
\begin{equation}\label{exp}
A = R e^{i\theta}.
\end{equation}
Then the Eq. (~\ref{GLequ}) becomes
\begin{equation}
\begin{split}
&\partial_{t}R=[\bigtriangleup - (\bigtriangledown\theta)^2]R - b(2\bigtriangledown\theta\cdot\bigtriangledown R 
+ R\bigtriangleup\theta) + (1 - R^2)R,\\
&R\partial_{t}\theta = b[\bigtriangleup - (\bigtriangledown\theta)^2]R 
+ 2\bigtriangledown\theta\cdot\bigtriangledown R + R\bigtriangleup\theta -cR^3.
\end{split}
\end{equation}
If $b=0$, then it becomes a class of reaction-diffusion equations called $\lambda - \omega$ systems
which have the following general form
\begin{equation}
\begin{split}
&\partial_{t}R=[\bigtriangleup - (\bigtriangledown\theta)^2]R  + R\lambda(R),\\
&R\partial_{t}\theta= 2\bigtriangledown\theta\cdot\bigtriangledown R + R\bigtriangleup\theta + qR\omega(R).\label{rw}
\end{split}
\end{equation}
For one-dimensional systems, let us assume the analogues of spiral wave 
solutions have the form 
\begin{equation}
\begin{split}
&R = \rho(x), \\
&\theta = -c(1 - k^2)T + \psi(x),
\end{split}
\end{equation}
where $k$ is an arbitrary constant and $T$ is a long-time scale \cite{hagan1982}. 
Substitution into Eq. (~\ref{rw}) yields~\cite{hagan1982}
\begin{equation}
\begin{split}
&\rho_{xx} + \rho(1 - \rho^2 - \psi_{x}^2) = 0,\\
&\psi_{xx} + 2\rho_{x}\psi_{x}/\rho = -c(1 - k^2 - \rho^2).
\end{split}
\end{equation}
Let us assume,
\begin{equation}
\begin{split}
&\xi=fx,\\
&\rho = a_0 + a_1F(\xi) + ... + a_mF(\xi)^m,\\
&\psi_x = b_0 + b_1F(\xi) + ... + b_nF(\xi)^n.
\end{split}
\end{equation}
where $a_i$ and $b_j$ ($i = 0, 1, m; j = 0, 1, ...n$) are constants to be 
determined later. $F(\xi)$ is a solution of the first-oder nonlinear ODE
\begin{equation}
\begin{split}
&F^{'2} = s_4F^4 + s_2F^2 + s_0,\\
&f^2\rho_{xx} + \rho(1 - \rho^2 - \psi_{x}^2) = 0,\\
&f\psi_{xx} + 2f\rho_{x}\psi_{x}/\rho = -c(1 - k^2 - \rho^2).
\end{split}
\end{equation}
According to the homogeneous balance method~\cite{yangkq2004}, we know $m = n = 1$, hence
\begin{equation}
\begin{split}
\rho = a_0 + a_1F(\xi),\\
\psi_x = b_0 + b_1F(\xi).
\end{split}
\end{equation}
So we get the equations,
\begin{equation}\label{GL1}
\begin{split}
&a_1f^2F^{''} + (a_0 + a_1F) - (a_0 + a_1F)^3 - (a_0 + a_1F)(b_0 + b_1F^2)^2 = 0,\\
&(a_0 + a_1F)b_1fF^{'} + 2a_1fF^{'}(b_0 + b_1F^2) = -c(1 - k^2)(a_0 + a_1F) + c(a_0 + a_1F)^3,\\
&2a_1s_4f^2F^3 + a_1s_2f^2F + a_0 + a_1F -a_0^3 - 3a_0^2a_1F - 3a_0a_1^2F^2 - a_1^3F^3\\
&- a_0b_0^2 - 2a_0b_0b_1F - a_0b_1^2F^2 - a_1b_0^2F - 2a_1b_0b_1F^2 - a_1b_1^2F^3=0.
\end{split}
\end{equation}
By requiring the coefficients of each term $F^i$ in the third equation 
of Eqs. (\ref{GL1}) are zero, we obtain,
\begin{equation}
\begin{split}
&F^3: 2a_1s_4f^2-a_1^3-a_1b_1^2=0,\\
&F^2: -3a_0a_1^2-a_0b_1^2-2a_1b_0b_1=0,\\
&F^1: a_1s_2f^2+a_1-3a_0^2a_1-2a_0b_0b_1-a_1b_0^2=0,\\
&F^0: a_0-a_0^3-a_0b_0^2=0.
\end{split}
\end{equation}
From the above equations we obtain
$a_0=b_0=0$, $f^2=-1/s_2$, $a_1^2+b_1^2=2s_4f^2$.
Then from the first two equations of Eqs. (~\ref{GL1}) we have
\begin{equation}
3b_1fF^{'} = -c(1 - k^2) - a_1^2qF^2.
\end{equation}
Finally, we obtain
\begin{equation}\label{s0s2s4}
\begin{split}
&4s_0s_4 = s_2^2, s_2 <0,\\
&a_1^2 = -s_2(1-k^2)/(2s_0),\\
&b_1^2 = -c^2(1 - k^2)^2s_2/(9s_0).
\end{split}
\end{equation}

% This is Table 1
\begin{table}[tbh]
\begin{center}
\begin{tabular}{ccccc} \hline
$s_4$ & $s_2$ & $s_0$ & $F^{'2} = s_4F^4 + s_2F^2 + s_0$ & $F(x)$ \\ \hline
$m^2$ & $-(1+m^2)$ & $1$ & $F^{'2} = (1-F^2)(1-m^2F^2)$ & $sn(x)$ \\
$-m^2$ & $2m^2-1$ & $1-m^2$ & $F^{'2} = (1-F^2)(1+m^2F^2-m^2)$ & $cn(x)$ \\
$-1$ & $2-m^2$ & $m^2-1$ & $F^{'2} = (1-F^2)(F^2+m^2-1)$ & $dn(x)$ \\
$1$ & $-(1+m^2)$ & $m^2$ & $F^{'2} = (1-F^2)(m^2-F^2)$ & $ns(x)$ \\
$1-m^2$ & $2m^2-1$ & $-m^2$ & $F^{'2} = (1-F^2)[(m^2-1)F^2-m^2]$ & $nc(x)$ \\
$m^2-1$ & $2-m^2$ & $-1$ & $F^{'2} = (1-F^2)[(1-m^2)F^2-1]$ & $nd(x)$ \\
$-m^2(1-m^2)$ & $2-m^2$ & $1$ & $F^{'2} = (1+F^2)[(1-m^2)F^2+1]$ & $sc(x)$ \\
$-m^2(1-m^2)$ & $2m^2-1$ & $1$ & $F^{'2} = (1+m^2F^2)[(m^2-1)F^2+1)$ & $sd(x)$\\
$1$ & $2-m^2$ & $1-m^2$ & $F^{'2} = (1+F^2)(1-m^2+F^2)$ & $cs(x)$ \\
$1$ & $2m^2-1$ & $-m^2(1-m^2)$ & $F^{'2} = (m^2+F^2)(m^2-1+F^2)$ & $ds(x)$ \\
\hline
\end{tabular}
\end{center}
\caption{\label{tab:paraF} Relations between the parameter$(s_0, s_2, s_4)$ and $F(x)$, where
$F(x)$ is satisfied with ODE $F^{'2} = s_4F^4 + s_2F^2 + s_0$. }
\end{table}
According to the Table \ref{tab:paraF}, the only possible Jacobi elliptic
function which is satisfied with the Eqs. (\ref{s0s2s4}) is $sn(\xi)$
with $m= 1$, i.e. $\tanh(\xi)$ ~\cite{yangkq2004}.
So we obtain,
\begin{equation}
\begin{split}
&a_0=b_0=0,\\
&a_1=\sqrt{1-k^2},\\
&b_1=k,\\
&F(\xi)= \tanh(\xi),\\
&f=1/\sqrt{2}.
\end{split}
\end{equation}
So the spiral wave solution for the one-dimensional GL Equation is
\begin{equation}
\begin{split}
&\rho(x)= \sqrt{1-k^2}\tanh(x/\sqrt{2}),\\
&\psi(x)= k \tanh(x/\sqrt{2}),\\
&c = -3k/(\sqrt{2}(1-k^2)).
\end{split}
\end{equation}
So we finally get the one-dimensional GL Equation's solution
\begin{equation}
A = \sqrt{1-k^2}\tanh(x/\sqrt{2})e^{i[3k(1-k^2)T/(\sqrt{2}(1-k^2)) + k \tanh(x/\sqrt{2})]}.
\end{equation}
This is the same as Hagan's solution~\cite{hagan1982}.

\section{Conclusions}
Based on balance mechanism in nonlinear PDEs, the extended F-expansion method
is widely used to obtain single and combined non-degenerative Jacobi elliptic
function solutions, as well as their corresponding degenerative solutions, for 
many kinds of PDEs. As example, we give a spiral wave solution for the one-dimensional
GL equation in this paper. From this example we can see that the extended F-expansion method
is a powerful tool in the nonlinear PDE field. We can expect that solutions of 
the higher dimensional GL Equations also will be obtained in this way.

\end{document}